\documentstyle[11pt]{article}

\global\arraycolsep=1pt 
\begin{document}

\font\cmss=cmss10 \font\cmsss=cmss10 at 7pt \hfill HUTP-96/A037 
 
\hfill MIT-CTP-2560 
 
\hfill BRX-TH-396 
 
\hfill hepth/9608125 
 
\hfill August, 1996 
 
\vspace{4pt} 
 
\begin{center} 
{\large {\bf \vspace{10pt} UNIVERSALITY OF THE OPERATOR PRODUCT EXPANSIONS \\%
[0pt] 
OF SCFT$_4$}} 
 
\vspace{10pt} 
 
{\sl D. Anselmi}$^{a}${\sl , D.Z. Freedman}$^{b}${\sl , M.T. Grisaru}$^{c}$%
{\sl , A.A. Johansen}$^{a}$ 
\end{center} 
 
\vspace{4pt} 
 
$^{a}${\it Lyman Laboratory, Harvard University, Cambridge, MA 02138, USA} 
 
$^{b}${\it Department of Mathematics and Center for Theoretical Physics, 
Massachussetts Institute of Technology, Cambridge MA 02139, USA} 
 
$^{c}${\it Physics Department, Brandeis University, Waltham MA\ 02254, USA} 
 
\vspace{12pt} 
 
\begin{center} 
{\bf Abstract} 
\end{center} 
 
\vspace{4pt} We study the operator product algebra of the supercurrent $%
J_{\alpha \dot{\alpha}}$ and Konishi superfield $K$ in four-dimensional 
supersymmetric gauge theories. The Konishi superfield appears in the $JJ$ 
OPE and the algebra is characterized by two central charges $c$ and $%
c^{\prime }$ and an anomalous dimension $h$ for $K$. In free field 
(one-loop) approximation, $c\sim 3N_{v}+N_{\chi }$ and $c^{\prime }\sim 
N_{\chi }$, where $N_{v}$ and $N_{\chi }$ are, respectively, the number of 
vector and chiral multiplets in the theory. In higher order $c$, $c^{\prime 
} $ and $h$ depend on the gauge and Yukawa couplings and we obtain the 
two-loop contributions by combining earlier work on $c$ with our own 
calculations of $c^{\prime }$ and $h$. The major result is that the radiative 
corrections to the central charges cancel when the one-loop beta-functions 
vanish, suggesting that $c$ and $c^{\prime }$ are invariant under continuous 
deformations of superconformal theories. The behavior of $c$ and $c^{\prime 
} $ along renormalization group flows is studied from the viewpoint of a $c$%
-theorem. \vfill\eject 
 
\setcounter{equation}{0}\vspace{0.2in} 
 
{\em Introduction.\vspace{0.05in}} 
 
In the past few years the study of superconformal quantum field theories in 
four spacetime dimensions (SCFT$_{4}$) has received renewed attention.  
Examples of superconformal invariant theories such as N=4 super Yang-Mills 
theory and certain N=2 and N=1 supersymmetric gauge theories with suitably 
chosen matter content \cite{finite} have been known for some time. 
Asymptotically free supersymmetric gauge theories can flow to interacting 
superconformal theories in the infrared and the infrared fixed points can 
contain important information about nonperturbative dynamics \cite{Seiberg}. 
Furthermore, there have been studies of the renormalization group flows from 
the viewpoint of a possible $c$-theorem \cite{zamolo} in four dimensions  
\cite{ctheorem,capp}. 
 
In two-dimensional conformal field theories the operator products of the 
stress tensor and conserved currents reflect the quantum properties of the 
conformal symmetry \cite{polyzamolo}. A study of the analogous OPE's in four 
dimensional N=1 supersymmetric theories was initiated in \cite{famous} and a 
very striking difference with respect to the two-dimensional case was 
emphasized. One considers the supercurrent superfield $J_{\alpha \dot{\alpha}%
}(z)$ ($z=x,~\theta ,~\bar{\theta}$) which contains the stress tensor $%
T_{\mu \nu }(x)$, the supercurrent $S_{\mu }(x)$ and the R-symmetry current $%
R_{\mu }(x)$ as its physical components. The lowest dimensional operator 
that appears in the $J_{\alpha \dot{\alpha}}(z)\hspace{0.03in}J_{\beta \dot{%
\beta}}(z^{\prime })$ OPE is a scalar superfield $\Sigma (z)$ of dimension $%
2+h$, where $h$ is an anomalous dimension depending on the coupling 
constants of the theory. As we discuss in more detail below, $\Sigma (z)$ is 
related to the Konishi superfield $K(z)=\bar{\Phi}{\rm e}^{V}\Phi $ ($\Phi $ 
and $V$ are the chiral and gauge superfields of the theory), whose physical 
components are $K(x)=\bar{\phi}(x)\phi (x)$, an axial current $K_{\mu }(x)=%
\frac{1}{2}\bar{\psi}\gamma _{\mu }\gamma _{5}\psi +\bar{\phi}\stackrel{%
\leftrightarrow }{D}_{\mu }\phi $, and the matter kinetic Lagrangian 
density. In a schematic notation the OPE's are expected to take the form  
\cite{famous}  
\begin{eqnarray} 
J(z)J(z^{\prime }) &=&{\frac{c}{(s\bar{s})^{3}}}+{\frac{\Sigma (z^{\prime })%
}{(s\bar{s})^{2-h/2}}}+\cdots ,  \nonumber \\ 
J(z)\Sigma (z^{\prime }) &=&{\frac{h\Sigma (z^{\prime })}{(s\bar{s})^{3/2}}}%
+[D,\bar{D}\}\frac{\Sigma (z^{\prime })}{s\bar{s}}+\cdots ,  \label{ope} \\ 
\Sigma (z)\Sigma (z^{\prime }) &=&{\frac{c^{\prime }}{(s\bar{s})^{2+h}}}+%
\frac{\Sigma (z^{\prime })}{(s\bar{s})^{1+h/2}}+\cdots  \nonumber 
\end{eqnarray} 
where $s=x-x^{\prime }+\theta $-$terms$ is a (chiral) superspace interval 
and $D,\bar{D}$ are spinor derivatives. 
 
The three numbers $c$, $c^{\prime }$ and $h$ characterize a superconformal 
theory in four dimensions. The central charge $c$ is related to the 
gravitational trace anomaly of the theory. In a free field theory with $%
N_{1} $, $N_{1/2}$ and $N_{0}$ real or Majorana fields of spin 1, 1/2 and 0 
respectively, one-loop calculations give \cite{somebody}  
\begin{equation} 
c=\frac{1}{120}\left( 12N_{1}+3N_{1/2}+N_{0}\right) .  \label{c} 
\end{equation} 
In a supersymmetric gauge theory with $N_{v}={\rm \dim \hspace{0.03in}}G$ 
vector multiplets, $G$ denoting the gauge group, and $N_{\chi }={\rm \dim  
\hspace{0.03in}}T$ chiral multiplets in the representation $T$, one can 
rewrite this as  
\begin{equation} 
c=\frac{1}{24}\left( 3N_{v}+N_{\chi }\right) .  \label{csusy} 
\end{equation} 
In a free field theory the operator $\Sigma (z)$ is just $\bar{\Phi}%
\Phi $ and one has $h=0$ and \cite{famous}  
\begin{equation} 
c^{\prime }=N_{\chi },  \label{c'} 
\end{equation} 
Thus $c$ is a measure of the total number of degrees of freedom of the 
theory, while $c^{\prime }$ measures the number of chiral matter multiplets.%
\vspace{0.2in} 
 
{\em Quantum corrections.\vspace{0.05in}} 
 
We will study lowest order quantum corrections to $c$, $c^{\prime }$ and $h$%
, in interacting N=1 supersymmetric gauge theories with gauge group $G$ and 
gauge coupling $g$. Chiral superfields $\Phi ^{i}$ are assigned to the 
representation $T^{a}$ of $G$, in general reducible, and there is a cubic 
superpotential $W=\frac{1}{6}Y_{ijk}\Phi ^{i}\Phi ^{j}\Phi ^{k}$. Most of 
our computations were carried out in components of the vector multiplet (in 
Wess-Zumino gauge) $V^{a}(z)\rightarrow A_{\mu }^{a}(x),~\lambda ^{a}(x),$ 
and the scalar multiplet $\Phi ^{i}(z)\rightarrow \phi ^{i}(x),~\psi ^{i}(x), 
$ with auxiliary fields eliminated. Using the formalism of Euclidean 
Majorana spinors of Nicolai \cite{Nicolai} the action reads  
\begin{eqnarray} 
S &=&\int d^{4}x\left[ \frac{1}{4}F_{\mu \nu }^{~~2}+\frac{1}{2}\bar{\lambda}%
{D}\!\!\!\!\slash\lambda +\overline{D_{\mu }\phi }D_{\mu }\phi +\frac{1}{2}%
\bar{\psi}{D}\!\!\!\!\slash\psi \right.   \nonumber \\ 
&&+i\sqrt{2}g(\bar{\lambda}^{a}\bar{\phi}_{i}T_{~~j}^{ai}L\psi ^{j}-\bar{\psi%
}_{i}RT_{~~j}^{ai}\phi ^{j}\lambda ^{a})  \nonumber \\ 
&&-\frac{1}{2}(\bar{\psi}^{i}LY_{ijk}\phi ^{k}\psi ^{j}+\bar{\psi}_{i}R\bar{Y%
}^{ijk}\bar{\phi}_{k}\psi _{j})  \nonumber \\ 
&&\left. +\frac{1}{2}g^{2}(\bar{\phi}_{i}T_{~~j}^{ai}\phi ^{j})^{2}+\frac{1}{%
4}Y_{ijk}\bar{Y}^{ilm}\phi ^{j}\phi ^{k}\bar{\phi}_{l}\bar{\phi}_{m}\right]  
\label{lagra} 
\end{eqnarray} 
where $L$ and $R$ are projection operators $L,R=\frac{1}{2}(1\mp \gamma _{5}) 
$. Note that the index $i$ is both a flavor and color index. 
 
The theory contains the classically conserved, but anomalous, $R$-current  
\begin{equation} 
R_{\mu }(x)=\frac{1}{2}\bar{\lambda}\gamma _{\mu }\gamma _{5}\lambda -\frac{1%
}{6}\bar{\psi}\gamma _{\mu }\gamma _{5}\psi +\frac{2}{3}\bar{\phi}\stackrel{%
\leftrightarrow }{D}_{\mu }\phi  
\end{equation} 
which is the $\theta =0$ component of the superfield $J_{\alpha \dot{\alpha}} 
$. Assuming that the matter multiplets are in an irreducible representation of the gauge group,
the Konishi operator is the unique renormalizable non-chiral deformation 
of a critical supersymmetric gauge theory. 
Its axial vector component $%
K_{\mu }(x)=\frac{1}{2}\bar{\psi}_{i}\gamma _{\mu }\gamma _{5}\psi ^{i}+\bar{%
\phi}_{i}\stackrel{\leftrightarrow }{D}_{\mu }\phi ^{i}$ is not conserved 
classically if the theory contains a superpotential, and it also suffers the 
Konishi anomaly \cite{Konishi}. Both effects generate an anomalous dimension  
$h$, so its scale dimension is $2+h$.

In our two-loop approximation the 
anomalous dimension $h$ is only due to the superpotential.
We can give a simple argument for this.
As we shall see in detail,
a correlator like $\hbox{$<K_\mu(x)K_\nu(0)>$}$ 
maintains its conformal properties
to the two-loop order. Conformal symmetry requires that it
is proportional to
$\hbox{$(\delta_{\mu\nu}-2x_\mu x_\nu/|x|^2)/|x|^{6+2h}$}$, so 
its divergence $<\partial K(x)K_\nu(0)>$ is
proportional to $h$. On the other hand $\partial K$  
is the sum of two contributions, from the
anomaly, proportional to $F \tilde F$, 
and from the superpotential. However, the 
former only starts contributing at three-loop order, as can be seen by
considering the relevant Feynman diagrams. 
Hence, only the superpotential
term contributes to $h$. For a similar reason
$ R_{\mu }$ has no anomalous dimension to our order.

The free field calculations of \cite{famous} revealed the presence of the 
operator $\Sigma_{free}(z)=\bar{\Phi}_{i}\Phi ^{i}$ in the $J_{\alpha \dot{\alpha}%
}(z)J_{\beta \dot{\beta}}(z^{\prime })$ OPE in (\ref{ope}). In the 
interacting case we {\em define} (see eq. (\ref{opeca})) the operator which 
appears in this position as the real superfield $\Sigma (z)$ and refer to 
its lowest component as $\Sigma (x)$. We can write $\Sigma (z)=\rho (g,Y)K(z)$, since $K(z)$ is the only gauge invariant composite operator
of the correct dimension.  
$\rho (g,Y)$ is a function of the couplings that we determine in lowest 
nontrivial (two-loop) order. $K$ is the renormalized 
Konishi operator defined to carry the power $\mu^h$ of the renormalization 
scale, so that $\rho$ is dimensionless. 
 
For simplicity our discussion has assumed that matter multiplets are
in an irreducible representation of the gauge group. In the reducible
case there is a Konishi superfield $K_I$ for each irreducible component,
and the $K_I$ mix under renormalization. Nevertheless, as will be shown
elsewhere \cite{97}, there is a unique central charge
associated with the set of $K_I$ which coincides through two-loop order
with $c'$ computed below.

Quantum corrections to $c$ have been computed previously \cite{jack}, and we 
shall present them below. To compute corrections to $c^{\prime }$ and $h$ it 
proved most convenient to study the OPE's (\ref{ope}) using the $R_{\mu }(x)$ 
component of $J_{\alpha \dot{\alpha}}(z)$ and the $K(x)$ (and also the $%
K_{\mu }(x)$) components of $K(z)$. 
 
The component OPE's  
\begin{eqnarray} 
\left. R_{\mu }(x)R_{\nu }(y)\right| _{x\rightarrow y}=_{{}} &&\frac{1}{3\pi 
^{4}}(\partial _{\mu }\partial _{\nu }-\Box \delta _{\mu \nu })\frac{c}{%
(x-y)^{4}}  \nonumber \\ 
&&+\frac{2}{9\pi ^{2}}\Sigma (y)(\partial _{\mu }\partial _{\nu }-\delta 
_{\mu \nu }\Box )\frac{1}{(x-y)^{2-h}}+\cdots  \label{opeca} \\ 
\left. \Sigma (x)\Sigma (y)\right| _{x\rightarrow y}=_{_{{}}} &&\frac{1}{%
16\pi ^{4}}\frac{c^{\prime }}{(x-y)^{4+2h}}+\cdots  \label{opec} 
\end{eqnarray} 
where $+\cdots $ denote less singular terms, precisely define the central 
charges $c$, $c^{\prime }$, the anomalous dimension $h$, and the operator $%
\Sigma (x)$. The numerical factor $1/3\pi ^{4}$ was chosen so that $c$ has 
the same value in the $R_{\mu }R_{\nu }$ OPE as in its conventional 
definition (\ref{c}) from the $T_{\mu \nu }T_{\rho \sigma }$ OPE and the 
curved space trace anomaly. The transverse tensor structure in (\ref{opeca}) 
is valid to two-loop order, even though $R_{\mu }$ is not conserved off 
criticality due to the chiral anomaly. 
To prove this, it is sufficient to consider the divergence
$\left.\partial R(x)\,R_\nu(y)\right| _{x\rightarrow y}$. 
The anomaly operator equation shows that this OPE
is proportional to $g^2 F\tilde F(x)\, R_\nu(y)$, which
does not contribute to the right hand side of (\ref{opeca}).
The effects of nonconservation appear 
begining at three-loop order where additional 
nonconserved tensor structures 
are present. 
 
We have performed two independent computations of $c^{\prime }$ and $h$. The 
first one involved the study of the connected four-point correlation 
function $<R_{\mu }(x)R_{\nu }(y)R_{\rho }(z)R_{\sigma }(w)>$ in the 
asymptotic region where $|x-y|,|z-w|\ll |x-z|,|y-w|$. In this limit we 
expect the $R_{\mu }(x)R_{\nu }(y)$ and $R_{\rho }(z)R_{\sigma }(w)$ OPE's 
to be dominated by the operator $\Sigma (y)$ so that, according to (\ref 
{opeca})-(\ref{opec}), we should find the asymptotic form  
\begin{eqnarray} 
&<&R_{\mu }(x)R_{\nu }(y)R_{\rho }(z)R_{\sigma }(w)>\sim \frac{c^{\prime }}{%
324\pi ^{8}}(\partial _{\mu }\partial _{\nu }-\delta _{\mu \nu }\Box )\frac{1%
}{(x-y)^{2-h}}  \nonumber \\ 
&&\times \frac{1}{(x-z)^{4+2h}}\hspace{0.03in}(\partial _{\rho }\partial 
_{\sigma }-\delta _{\rho \sigma }\Box )\frac{1}{(z-w)^{2-h}}.  \label{asy} 
\end{eqnarray} 
Note that no scale appears in (\ref{asy}) which is what is expected from a 
correlator of conserved currents of fixed dimension 3 in a conformal theory. 
 
In perturbation theory, the effect of $h$ is logarithmic, so that we 
computed the asymptotic form of all one- and two-loop four-point graphs 
looking for terms of the form  
\begin{eqnarray} 
<R(x)R(y)R(z)R(w)> &\sim &\frac{1}{(x-y)^{4}(x-z)^{4}(z-w)^{4}}  \nonumber \\ 
&.&\left[ N_{\chi }+ag^{2}+a^{\prime }|Y|^{2}+a^{\prime \prime }|Y|^{2}\ln  
\frac{|x-y||z-w|}{|x-z|^{2}}\right]  
\end{eqnarray} 
Graphs which contain the indicated asymptotic structure all have $\phi \bar{%
\phi}$ lines in the intermediate state in the $x-z$ channel, as expected 
from the relation between $\Sigma (x)$ and $K(x)$. For the superpotential 
contribution, to avoid some technical problems due to the anomalous 
dimension of the Konishi operator, we computed the four-point correlator of $%
J_{\alpha \dot{\alpha}}$ in superspace. The gauge interactions do not 
contribute to the anomalous dimension and the simplest way to compute their 
contributions to the correlator (\ref{asy}) was to work in components using 
the Feynman gauge for gluons. This gauge is convenient, because the one-loop 
correction to the scalar propagator due to the virtual gluons and gluini 
vanishes. Regularization by dimensional reduction was used in the 
intermediate stage of the computation, although divergences cancel in the 
final contributions to $a$, $a^{\prime }$ and $a^{\prime \prime }$. One 
should note that this method of calculation does not give the 
proportionality factor $\rho (g,Y)$ between $\Sigma (z)$ and $K(z)$. 
 
Our second method requires the two-loop computation of the two- and 
three-point correlators $<K(x)K(y)>$ and $<R_{\mu }(x)R_{\nu }(y)K(z)>$. We 
exploit the well-known fact that in a conformally invariant theory, the 
conformal symmetry largely, indeed frequently uniquely, fixes the form of 
two- and three-point correlation functions \cite{Osborn}. 
The methods of conformal symmetry are applicable 
here despite the usual lore that the divergences of perturbation theory 
spoil conformal properties of a theory. Briefly, the reasons are that 
correlation functions for non-coincident spatial points have no overall 
divergence, and the only subdivergences which lead to violation of conformal 
Ward identities correspond to renormalization of the couplings $g$ and $Y$. 
Such renormalization subdivergences do not appear until three-loop order in 
the correlators $<R_{\mu }(x)R_{\nu }(y)K(z)>$ and $<K(x)K(y)>$. 
 
Scale and translation properties are sufficient to determine the form  
\begin{equation} 
<K(x)K(y)>=\frac{A}{16\pi ^{4}(x-y)^{4+2h}}  \label{G} 
\end{equation} 
with $A=c^{\prime }/\rho ^{2}$. The tensor form of $<R_{\mu }(x)R_{\nu 
}(y)K(z)>$ can be fixed by the additional requirements of conservation and 
correct transformation properties under the discrete inversion element of 
the conformal group, i.e. $x^{\mu }\rightarrow x^{\prime }{}^{\mu }=x^{\mu 
}/x^{2}$. The result is the conformal tensor  
\begin{equation} 
<R_{\mu }(x)R_{\nu }(y)K(z)>=\frac{B}{36\pi ^{6}}\frac{\left( 1-\frac{h}{4}%
\right) J_{\mu \nu }(x-y)-\frac{1}{2}\left( 1+\frac{h}{2}\right) J_{\mu \rho 
}(x-z)J_{\rho \nu }(z-y)}{(x-y)^{4-h}(x-z)^{2+h}(y-z)^{2+h}},  \label{H} 
\end{equation} 
where the tensor $J_{\mu \nu }(x)=\delta _{\mu \nu }-2\hspace{0.02in}x_{\mu 
}x_{\nu }/x^{2}$ is an orthogonal matrix which is (essentially) the Jacobian 
of the coordinate inversion  
\cite{Schreier,BakerJohnson,Osborn}. In the limit $x\sim y$ one finds the 
most singular term  
\begin{equation} 
<R_{\mu }(x)R_{\nu }(y)K(z)>\sim \frac{B}{72\pi ^{6}(h-2)}{\frac{1}{%
(y-z)^{4+2h}}}(\partial _{\mu }\partial _{\nu }-\Box \delta _{\mu \nu })%
\frac{1}{(x-y)^{2-h}}. 
\end{equation} 
Use of the OPE's (\ref{opeca})-(\ref{opec}) then immediately yields $%
c^{\prime }/\rho =B/(h-2)$. Then, a computation of the indicated two- and 
three-point correlators gives $A$, $B$ and $h$, and determines  
\begin{equation} 
c^{\prime }(g,Y)=\frac{B^{2}}{(h-2)^{2}A}\hspace{0.2in},\hspace{0.3in}\rho =%
\frac{B}{(h-2)A}.  \label{cprimo} 
\end{equation} 
The first expression allows us to check
that $c^{\prime }$ does not depend on the 
subtraction scheme. Indeed, scheme dependence manifests itself as the 
arbitrariness of redefining the renormalized operator $K$ with a finite\ 
multiplicative factor of the form $1+a|Y|^{2}$. Such a redefinition affects $%
\rho $, as well as $A$ and $B$, but cancels in the formula for $c^{\prime }$ 
and in the expression $\Sigma =\rho K$. 
 
We used conformal symmetry methods \cite{BakerJohnson} to facilitate the 
calculation of the Feynman diagrams. In general these methods are quite 
straightforward for graphs with virtual scalars and spinors and more 
difficult for virtual photons \cite{BakerJohnson} because the photon 
propagator transforms covariantly under a conformal transformation only when 
accompanied by a gauge transformation. For this reason we studied separately 
the contribution to $c^{\prime }$ and $\rho $ from the superpotential and 
from the gauge interactions of the Lagrangian (\ref{lagra}). In the 
superpotential sector we simply combined conformal methods and those of 
differential renormalization \cite{diffren} to give an exact calculation of 
all two-loop diagrams for the correlators and calculated $c^{\prime }$ from 
the quotient (\ref{cprimo}). 
 
For the gauge sector, in order to take advantage of existing calculations, 
we considered correlators of the axial 
vector component of $K(z)$ rather than its scalar component. A discussion 
similar to that given above applies to the correlators $%
\hbox{$<R_{\mu }(x)R_{\nu 
}(y)K_{\rho }(z)>$}$ and $<K_{\mu }(x)K_{\nu }(y)>$. There are unique 
conformal invariant tensor forms \cite{Schreier} whose coefficients are the 
same as $A$ and $B$ in (\ref{G}) and (\ref{H}) because of supersymmetry. The 
two-loop graphs contributing to $<R_{\mu }(x)R_{\nu }(y)K_{\rho }(z)>$ lead 
to integrals that have been evaluated by conformal techniques in previous 
work on possible radiative corrections to the chiral anomaly in quantum 
electrodynamics \cite{BakerJohnson} and the standard model \cite 
{EhrlichFreedman}. In all these cases the net two-loop contributions to the 
axial three-point function vanishes, which means that $B$ in the gauge 
sector of our problem is exactly given by the one-loop value. This 
computation did not require any regularization because we used a gauge in 
which the required vertex and self-energy insertions were finite. The 
correlator $<K_{\mu }(x)K_{\nu }(y)>$ and the value of A was then obtained 
from a superspace computation of $<K(z)K(z^{\prime })>$ using both 
dimensional reduction and differential renormalization which both gave the 
same result for $c^{\prime }$ in agreement with the first method. We do not 
give the result for $\rho $, for which the order $|Y|^{2}$ term is scheme 
dependent, because we have not described the scheme precisely enough for 
this to be meaningful.. The agreement of the two methods of calculation based 
on the four-point and on the combined three- and two-point correlators 
provides an excellent cross-check on the fundamental central charge $%
c^{\prime }$. 
 
The calculations described above give the results  
\begin{equation} 
c^{\prime}=N_{\chi}+2 \gamma^i_{~i}~~~~~,
~~~~~~~~h=\frac{3Y_{ijk}
\bar{Y}^{ijk}}{16\pi^2 N_{\chi}}%
 \label{op}
\end{equation} 
Quantum corrections to $c$ up to two loops are known from earlier work \cite 
{jack} for a general renormalizable theory of vectors, spinors and scalars 
in curved space. The results of \cite{jack} can be restated for a 
supersymmetric gauge theory as  
\begin{equation} 
c=\frac{1}{24}\left( 3N_{v}+N_{\chi }+N_{v}\frac{\beta (g)}{g}%
-\gamma_{{}}^{i}\hspace{0in}_{i }\right).  \label{c2loop} 
\end{equation} 
These results involve the gauge $\beta$-function and the anomalous dimension  
$\gamma^i_{~j}$ of the chiral superfields which, at the one-loop level are  
\cite{JackJones}  
\begin{eqnarray} 
16\pi ^{2}\hspace{0.02in}\beta(g) &=&\stackrel{}{g^{3}\left( -3C(G)+\frac{%
{\rm Tr\hspace{0.03in}}C(T)}{{\rm \dim \hspace{0.03in}}G}\right) },  
\nonumber \\ 
\beta _{ijk } &=&Y_{m (ij }\hspace{0.02in}\gamma _{{}}^{m }\hspace{0in}_{k 
)},  \nonumber \\ 
16\pi ^{2}\hspace{0.02in}\gamma _{{}}^{i }\hspace{0in}_{j } &=&\frac{1}{2}%
Y_{j k m }\bar Y^{i k m}-2g^{2}C(T)_{{}}^{i }\hspace{0in}_{j },  \label{betas} \\ 
C(G)\hspace{0.02in}\delta ^{ab} &=&f^{acd}f^{bcd},  \nonumber \\ 
C(T)_{{}}^{i }\hspace{0in}_{j } &=&(T^{a}T^{a})_{{}}^{i }\hspace{0in}_{j }.  
\nonumber 
\end{eqnarray} 
 
We observe now that the two-loop corrections to $c$ and
$c^{\prime }$ vanish if the conditions $\beta (g)=0$ and $\gamma
^i_{~i}=0$ are satisfied, and these conditions mean that the
theory is conformal invariant to one-loop
order. Conversely, if both $c$ and $c^{\prime }$ are uncorrected (not just $%
c $), then the theory is conformal. However, even under these conditions, 
for nonvanishing superpotential the anomalous dimension $h$ is not zero,
even for such superconformal theories as N=4 Yang-Mills theory. 

It is known that to all orders in the couplings $g$, $Y$ there is
a fixed surface of the renormalization group flow provided that
the matter
representation is chosen so that $3\hspace{0.02in}\dim {\rm \hspace{0.03in}}G%
\hspace{0.03in}C(G)-{\rm Tr\hspace{0.03in}}C(T)=0$ and $g$, $Y$ are such 
that 
$\gamma _{~j}^{i}=0$. This can be seen from the NSVZ relation \cite{nsvz}  
\[ 
16\pi ^{2}\hspace{0.02in}\beta ^{NSVZ}=-g^{3}{\frac{3C(G)\,{\rm \dim G}-%
{\rm Tr\hspace{0.03in}}[(1-2\hspace{0.02in}\gamma )\hspace{0.02in}C(T)]}{%
{\rm \dim G}\left( 1-\frac{g^{2}}{8\pi ^{2}}C(G)\right) }},  
\] 
and the relation between $\beta_{ijk}$ and the anomalous
dimension.  Therefore, there exists a space of continuously
connected conformal
invariant theories and we have found evidence that the two central charges $%
c $ and $c^{\prime }$ are universal, i.e. that $c$ and $c^{\prime
  }$ are constant, independent of the couplings, on this space.
We call such quantities SCFT$_4$ invariants. This universality
property is the four-dimensional analogue of the well known fact
that the Virasoro central charge of a two-dimensional conformal
theory is invariant under marginal deformations. The other
important information that we have obtained is that in four
dimensions the OPE's do contain a third quantity, namely $h$,
that is not an SCFT$_{4}$ invariant. $h$ tells us that
continuously connected SCFT$_{4}$'s are indeed inequivalent.
These results were anticipated in \cite{famous} and we regard
them as the essential features of superconformal (and conformal)
field theories in four dimensions.\vspace{0.2in}

{\em Possible }$c${\em -theorems.\vspace{0.05in}} 
 
At this point we discuss the search for a $c$-function in four-dimensional 
quantum field theory, a function of the couplings of the theory which must 
decrease along renormalization group flows toward the infrared reflecting 
the continouous ''integrating out'' of degrees of freedom and which is 
stationary at fixed points of the flow. Since there are two well-defined 
central charges in SCFT$_{4}$, we can consider $c$ and $c^{\prime }$, which 
according to (\ref{csusy}) and (\ref{c'}) measure, respectively, the total 
number of degrees of freedom and the total number of scalar multiplets. Or 
we can examine the quantity $N_{v}^{eff}=8c-c^{\prime }/3$ which is a 
measure of the effective number of vector multiplets. A $c$-theorem for one 
or more of these quantities $c$, $N_{v}^{eff}$ or $c^{\prime }$, could allow 
a test of phenomena such as confinement, chiral and supersymmetry breaking 
and the Higgs mechanism. There are known non-supersymmetric examples \cite 
{capp} of both increasing and decreasing behavior of $c$ along RG flows, so 
that $c$ is not a good $c$-function, but the situation could be better in 
supersymmetry \cite{bastianelli}. Our analysis will show that none of the 
quantities $c$, $c^{\prime }$ and $N_{v}^{eff}$ satisfies the desiderata of 
a $c$-function for the full class of N=1 supersymmetric gauge theories with 
superpotential. Yet, $c^{\prime }$ seems to have an interesting behaviour 
for the subclass of asymptotically free theories. 
 
We begin by examining theories with vanishing superpotential ($Y=0$). In 
this case we obtain from (\ref{c2loop})  
\begin{equation} 
c=\frac{1}{24}\left[ 3N_{v}+N_{\chi }+{\frac{3g^{2}}{16\pi ^{2}}}\left( 
-C(G)N_{v}+{\rm Tr\hspace{0.03in}}C(T)\right) \right] .  \label{cc} 
\end{equation} 
{} From (\ref{betas}) we learn that the theory is UV free if ${\rm Tr\hspace{%
0.03in}}C(T)<3C(G)\hspace{0.02in}N_{v}$, so that we can consider the change 
in $g$ as we move down from the UV fixed point. We see that $c$ decreases if  
${\rm Tr\hspace{0.03in}}C(T)<C(G)\hspace{0.02in}N_{v}$, but increases 
outside of this range. For SU(N$_{c}$) supersymmetric QCD with $N_{f}$ 
fundamental quarks and their anti- quarks, this equality becomes $N_{f}<N_{c} 
$ which is a small part of the range of asymptotic freedom $N_{f}<3N_{c}$ 
and, except for $N_{f}=0$, exactly the range for which non-perturbative 
considerations indicate that there is no ground state \cite{Seiberg2}. 
Remarks similar to those above hold for $N_{v}^{eff}$, but the range in 
which this quantity decreases, namely $N_{f}<3N_{c}/7$, is smaller than for $%
c$. So it is only for the case of pure supersymmetric Yang-Mills theory 
without chiral matter that we obtain a decreasing $c$-function. 
 
We can also restrict to supersymmetric gauge theories for which the one-loop  
$\beta $-function vanishes and examine the behavior of our $c$-function 
candidates on flows in the neighborhood of the fixed surface discussed 
above. This surface is known \cite{LeighStrassler} to be infrared attractive 
on both sides. But the radiative corrections to $c$, $c^{\prime }$ and $%
N_{v}^{eff}$ are linear in 
$\gamma _{~j}^{i}$, so that each of these quantities decreases along flows 
on one side of the fixed surface, but increases along flows on the other 
side. 
 
We now take a closer look at  
\begin{equation} 
c^{\prime}=N_\chi-{\frac{1}{4\pi^2}}\left[g^2 {\rm Tr}C(T)-{\frac{1}{4}}%
Y_{ijk} \bar Y^{ijk}\right]. 
\end{equation} 
We first note that when $Y=0$ $c^{\prime}$ decreases from the ultraviolet in 
an asymptotically free gauge theory. It is interesting to see in detail what 
happens when $g$ and $Y$ are both nonzero. The one-loop flow equations 
suggested by (\ref{betas}) read  
\begin{eqnarray} 
{\frac{dy}{dt}}&=&by(qy^2-rg^2),  \nonumber \\ 
{\frac{dg}{dt}}&=&-bg^3, 
\end{eqnarray} 
where $b>0$ in an asymptotically free theory. Dividing and writing $y=gz$ we 
obtain  
\begin{equation} 
g{\frac{dz}{dg}}=(r-1)z-qz^3. 
\end{equation} 
If the initial value of $z$ satisfies $z^2<{\frac{r-1}{q}}$, then there is a 
solution for $y(g)$ which behaves at small $g$ as $y\sim g^r$. So the Yukawa 
coupling can be asymptotically free in the presence of a non-abelian gauge 
coupling\footnotemark  
\footnotetext{ 
These conditions are satisfied in a simple model with $G=SU(N)$, a matter 
superfield $\phi^a(z)$ in the adjoint representation and superpotential 
involving the $SU(N)$ $d$-symbol, in which case $r=3$.}. If $r>1$ then $%
y\rightarrow 0$ faster than $g$ and $c^{\prime}$ is a decreasing function as 
we move away from the ultraviolet fixed point in the space of $g$, $Y$. 
 
The intriguing fact that $c^{\prime }$ appears to be a good $c$-function 
near the ultraviolet fixed point of asymptotically free gauge theories 
motivates us to test its infrared behavior by extending Bastianelli's 
analysis \cite{bastianelli} to $c^{\prime }$. The infrared fixed point of $%
SU(N_{c})$ supersymmetric gauge theory with $N_{c}+2\leq N_{f}<3N_{c}$ quark 
flavors is conjectured to be described by the magnetic dual theory with 
gauge group $SU(N_{f}-N_{c})$ with $N_{f}$ quarks, $N_{f}$ antiquarks and $%
N_{f}^{2}$ mesons. In the range $N_{c}+2\leq N_{f}<3N_{c}/2$ the magnetic 
theory is infrared free, so the free field values of $c$ and $c^{\prime }$ 
may be used. There are thus $c_{IR}^{\prime }=2N_{f}(N_{f}-N_{c})+N_{f}^{2}$ 
chiral multiplets in the magnetic theory and $c_{UV}^{\prime }=2N_{f}N_{c}$ 
in the electric theory. So one has $c_{UV}^{\prime }-c_{IR}^{\prime 
}=N_{f}(4N_{c}-3N_{f})$, which is positive only in part of the range for 
which the non-perturbative description in terms of a free dual theory is 
valid. Thus our attempted interpretation of $c^{\prime }$ as a $c$-function 
is problematic when the Seiberg scenario of non-perturbative dynamics is 
considered. 
 
It is worth stressing again the physical relevance of quantities that count 
the various kinds of degrees of freedom separately, in particular the 
quantity $N_{v}^{eff}$ that measures, in some sense, the size of the 
effective gauge symmetry group of the theory. Even if these quantities do 
not satisfy general $c$-theorems, it is still conceivable that they behave 
nicely in physical models. \vspace{0.2in} 
 
{\em Conclusions.} \vspace{0.05in} 
 
We summarize our results: for general supersymmetric theories,
the classical central charges $c$, $c^{\prime}$ and the dimension
of the operator $\Sigma$ receive quantum corrections. When the
conditions for superconformal invariance are satisfied the
corrections to the central charges vanish. This suggests that $c$
and $c^{\prime }$ are invariant under continuous deformations of
superconformal theories and generalizes the analogous property of
two-dimensional quantum field theory.  However, in four
dimensions the operator $\Sigma$ appears in the OPE's of the
supercurrent even at the conformal point, with nonvanishing
anomalous dimension $h$ whenever a superpotential (or Konishi
anomaly) is present.  Our N=1 component and superspace
calculations are also valid for extended N=2 and N=4 models,
because $c'$ and $h$ are scheme independent to the order
considered. In particular the stress-tensor OPE in the
conformally invariant N=4 theory contains components of the
Konishi operator with nonvanishing anomalous
dimension; extended supersymmetry does not simplify the form of
the OPE's.  We believe that these are the essential features of
the operator product algebra of superconformal field theories in
four dimensions. (In \cite{HoweWest} a form of the TT OPE for
  the $N=4$ theory is presented  containing only integer
  powers of the distance, which conflicts with our  result.)
 
We have also examined the possibility of establishing a 
$c$-theorem in terms of 
the quantity $c$ which counts the total number of multiplets, $c^{\prime }$ 
which counts the number of chiral multiplets, or the weighted difference $%
N_{v}^{eff}$ which counts the number of vector multiplets. We have found 
that none of them is a suitable candidate in general, although under some 
restrictive, and possibly physically relevant, conditions they might play 
such a role.\vspace{0.2in} 
 
{\em Acknowledments} \vspace{0.05in} 
 
Two of the authors (D.Z.F. and M.T.G.) are happy to thank the Aspen Center 
for Physics for its hospitality during the final phase of this work. 
Discussions with D.R.T. Jones, K. Johnson, M.J. Strassler and especially I. 
Jack were very helpful. The research of D.A. and A.A.J. was supported in 
part by the Packard Foundation and by NSF grant PHY-92-18167. The research 
of D.Z.F. was supported by NSF grant PHY-92-06867. The research of M.T.G. 
was supported by NSF grant PHY-92-22318. A.A.J was also supported by NATO 
grant GRG 93-0395.  
 

\vspace{0.2in} 

\begin{thebibliography}{99} 
\bibitem{finite}  For a recent discussion, see C. Lucchesi, All-order 
Finiteness in N=1 SYM Theories: Criteria and Applications, preprint 
NEIP-96-001, BUTP-96-7 and hep-ph/9604216. 
 
\bibitem{Seiberg}  N. Seiberg, Electric-Magnetic Duality in Supersymmetric 
Non-Abelian Gauge Theories, Nucl. Phys. B435 (1995) 129. 
 
\bibitem{zamolo}  A.B. Zamolodchikov, ``Irreversibility'' of the Flux of the 
Renormalization Group in a 2D Field Theory, JETP Lett. 43 (1986) 730. 
 
\bibitem{ctheorem}  J.L. Cardy, Is There a $c$-Theorem in Four Dimensions?, 
Phys. Lett B215 (1988) 749; 
 
H. Osborn, Derivation of a Four-Dimensional $c$-Theorem For Renormalizable 
Quantum Field Theories, Phys. Lett. B222 (1989) 97; 
 
I. Jack and H. Osborn, Analogs of the $c$-Theorem for Four-Dimensional 
Renormalizable Field Theories, Nucl. Phys. B343 (1990) 647; 
 
G. Shore, A New $c$-Theorem in Four-Dimensions, Phys. Lett. B253 (1991) 380; 
The $c_{{\cal F}}$ Theorem, {\it ibid.} B256 (1991) 407. 
 
\bibitem{capp}  A. Cappelli, D. Friedan and J.I. Latorre, $c$-Theorem and 
Spectral Representation, Nucl. Phys. B352 (1991) 616. 
 
\bibitem{polyzamolo}  A.A. Belavin, A.M. Polyakov and A.B. Zamolodchikov, 
Infinite Conformal Symmetry in Two-Dimensional Quantum Field Theory, Nucl. 
Phys. B241 (1984) 333. 
 
\bibitem{famous}  D. Anselmi, M. Grisaru and A.A. Johansen, A Critical 
Behaviour of Anomalous Currents, Electric-Magnetic Universality and CFT$_{4}$%
, HUTP-95/A048, BRX-TH-388 and hepth/9601023, January 1996. 
 
\bibitem{somebody}  S.M. Christensen and M.J. Duff, Axial and Conformal 
Anomalies for Arbitrary Spin in Gravity and Supergravity, Phys. Lett. 76B 
(1978) 571. 
 
\bibitem{Nicolai}  H. Nicolai, A Possible Constructive Approach to (Super-$%
\Phi ^{3}$)$_{4}$, Nucl. Phys. B140 (1978) 294. 
 
\bibitem{Konishi}  K. Konishi, Anomalous Symmetry Transformation of Some 
Composite Operators in SQCD, Phys. Lett. B135 (1984) 439. 
 
\bibitem{97} D. Anselmi, Central Functions and Their Physical Implications, 
to appear.

\bibitem{jack}  I. Jack, Background Field Calculations in Curved Spacetime, 
Nucl. Phys. B253 (1985) 323. 
 
\bibitem{Osborn}  
For a recent exposition of these ideas, see H.. Osborn and A. Petkou, Implications of Conformal 
Invariance in Field Theories for General Dimensions, Ann. Phys. (N.Y.) 231 
(1994) 311. 
 
\bibitem{Schreier}  E.J. Schreier, Conformal Symmetry and Three-Point 
Functions, Phys. Rev. D3 (1971) 980. 
 
\bibitem{BakerJohnson}  M. Baker and K. Johnson, Applications of Conformal 
Symmetry in Quantum Electrodynamics, Physica 96A (1979) 120. 
 
\bibitem{diffren}  D.Z. Freedman. K. Johnson and J.I. Latorre, Differential 
Regularization and Renormalization: a New Method of Calculation in Quantum 
Field Theory, Nucl. Phys. B371 (1992) 353. 
 
\bibitem{EhrlichFreedman}  J. Erlich and D.Z. Freedman,
Conformal Symmetry and the Chiral Anomaly, preprint
MIT-CPT-2588 and hepth/9611133. 
 
\bibitem{JackJones}  For a recent discussion, see I. Jack, D.R.T. Jones and 
C.G. North, $N=1$ supersymmetry and The Three Loop Gauge Beta Function,
Phys.Lett. B386 (1996) 13.
 
\bibitem{nsvz}  V. Novikov, M.A. Shifman, A.I. Vainshtein, V. Zakharov, 
Exact Gell-Mann-Low Function of Supersymmetric Yang-Mills Theories from 
Instanton Calculus, Nucl. Phys. B229 (1983) 381; 
 
M.A. Shifman and A.I. Vainshtein, Solution of the Anomaly Puzzle in 
supersymmetric Gauge Theories and the Wilson Operator Expansion, {\it ibid.} 
B277 (1986) 456. 
 
\bibitem{bastianelli}  F. Bastianelli, Tests for $c$-theorems in 4D, Phys. 
Lett B 369 (1996) 249. 
 
\bibitem{Seiberg2}  N. Seiberg, Exact Results on the Space of Vacua of Four 
Dimensional supersymmetric Gauge Theories, Phys. Rev. D 49 (1994) 6857. 
 
\bibitem{LeighStrassler}  R.G. Leigh and M.J. Strassler, Exactly Marginal 
Operators and Duality in Four Dimensional N=1 Supersymmetric Gauge Theory, 
Nucl. Phys. B447 (1995) 95. 
 
\bibitem{HoweWest}  P.S. Howe and P.C. West, Operator product expansions in 
four-dimensional superconformal field theories, preprint CERN-TH/96-159, 
King's College/kcl-th-96-13 and hepth/9607060. 
\end{thebibliography}
\end{document}